\documentclass[conference]{IEEEtran}
\IEEEoverridecommandlockouts
\usepackage{amsmath,amssymb,amsfonts}
\usepackage{algorithmic}
\usepackage{graphicx}
\usepackage{textcomp}
\usepackage{xcolor}
\usepackage{biblatex}

\allowdisplaybreaks
\usepackage{hyperref}

\begin{document}

\title{Timing configurations affect the macro-properties of multi-scale feedback systems\\
}

\author{
\IEEEauthorblockN{Patricia Mellodge}
\IEEEauthorblockA{\textit{Electrical \& Computer Engineering} \\
\textit{University of Hartford}\\
mellodge@hartford.edu}
\and
\IEEEauthorblockN{Ada Diaconescu}
\IEEEauthorblockA{\textit{Computer Science \& Networks} \\
\textit{Telecom Paris, LTCI}\\
diacones@telecom-paris.fr}
\and
\IEEEauthorblockN{Louisa Jane Di Felice}
\IEEEauthorblockA{\textit{Institute of Environmental Science \& Technology} \\
\textit{Autonomous University of Barcelona}\\
louisajane.difelice@uab.cat}
}

\maketitle

\begin{abstract}

Multi-scale feedback systems, where information cycles through micro- and macro-scales leading to adaptation, are ubiquitous across domains, from animal societies and human organisations to electric grids and neural networks. Studies on the effects of timing on system properties are often domain specific. The Multi-Scale Abstraction Feedbacks (MSAF) design pattern aims to generalise the description and understanding of multi-scale systems where feedback occurs across scales. We expand on MSAF to include timing considerations. We then apply these considerations to two models: a hierarchical oscillator (HO) and a hierarchical cellular automata (HCA). Results show how (i) different timing configurations significantly affect system macro-properties and (ii) different regions of time configurations can lead to the same macro-properties. These results contribute to theory, while also providing useful insights for designing and controlling such systems.

\end{abstract}


\section{Introduction}
\label{sec:intro}

Multi-scale systems are those systems where different scales of time, space, or information granularity interrelate via information flows. If information cycles through the system leading to the adaptation of system entities, they become \textit{multi-scale feedback systems} \cite{simon1991architecture}. For example, workers in an organisation send information about their state to their managers, who then send back commands leading to changes in their behaviours. Similarly, foraging ants lay pheromones, forming a trail that affects their behaviour. In autonomic  
systems, managed resources are monitored and control commands issued for self-adaptation \cite{kephart2003vision}. In these examples, information from the micro-scale (workers, ants, resources) is \textit{abstracted} onto a macro-scale, and some adaptation at the micro-scale occurs based on information flowing back down. Such feedback cycles can be repeated at recursively higher scales, with increasing abstraction tied to ever larger system parts, 
e.g., multi-level management organisations, autonomic systems \cite{kramer2007},\cite{weyns2013}, or plants `controlling' foraging ants to disperse their seeds, by attaching 
food packages to their grains \cite{ant-plants2007}. 
In this paper, we use both \textit{level} and \textit{scale} to express the idea of multiple amounts of time, space, or information granularity in a system.

 While multi-scale feedback systems can be found across all domains, their generic properties remain under-explored. In previous work, we introduced the \textit{Multi-Scale Abstraction Feedbacks} (MSAF) design pattern, as a means to generalise feedback cycles of information flows operating at multiple abstraction levels, in systems with different types of entities, structures, and functions \cite{diaconescu2019multi}\cite{diaconescu2021exogenous}. In this pattern, scales are identified in relation to 
 \textit{information abstraction}
and are orthogonal to how such abstractions are implemented. 
A macro-property at a higher scale can be tied to an \textit{exogenous} macro-entity (e.g., a manager in an organisation, different from the workers) but can also be \textit{micro-distributed} among micro-entities at a lower scale (e.g., knowledge of power relations distributed across members of an animal society \cite{flack2017coarse}), or \textit{composed} from the collective structure of micro-entities (e.g., forest patch shapes affecting tree growth \cite{filotas2014viewing}). Such multi-scale design allows coordinating 
increasingly large-scale systems, via a divide-and-conquer approach.  Each scale may process similar amounts of information by making a different trade-off between information accuracy and control scope. 

Another important trade-off is between a system's reaction time and the control scope considered, at different scales. Such timing aspects depend on inherent communication and processing delays, process execution frequencies, and adaptation lags (i.e., how long before adaptation takes effect). The question of how such 
timing configurations affect the behaviour of multi-scale systems, in particular, has been approached primarily in domain-specific ways. General 
insights that would facilitate cross-domain transfer remain vague and untested. 
E.g., it is often said that higher levels must operate 
slower than lower ones, to ensure 
system stability \cite{wu2013hierarchy}, \cite{kramer2007}, \cite{weyns2013}. Yet, excessive communication delays 
between macro- and micro-scales can 
cause dysfunction depending on system goals \cite{flack2013timescales}. 

This paper aims to reduce the gap between highly generic remarks and application-specific practices in matters of time concerns. 
First, we expand the MSAF pattern with domain-independent time-related aspects (sec.~\ref{MSAFpattern}).  
We then illustrate the impact of chosen timing aspects on system macro-properties via two generic case studies with multiple potential applications (secs.~\ref{biochemical} and \ref{HCA}). 
The studies expand previous multi-scale oscillator models: a biochemical model of hierarchical oscillators (HO) \cite{Kim2010} and a hierarchical cellular automata (HCA) \cite{diaconescu2018holonic}. We focus on these examples as many real-world systems are characterised by oscillating patterns, from collective behaviour in animal groups \cite{ermentrout1991adaptive} \cite{paley2007oscillator} and circadian cycles in the brain \cite{kang2009circadian} to opinion dynamics in social networks \cite{pluchino2006opinion}, clock synchronisation in distributed computing systems, 
and the coupled motion of pendulum clocks \cite{dilao2009antiphase}.

The contribution of our results is two-fold. First, we show how different timing configurations can significantly affect system macro-properties. This allows using time delays as configuration parameters for changing system behaviours. Second, we show how different regions of time configurations lead to the same macro-property. This can be used to improve system robustness to time disturbances. 
These general principles are relevant to domain practitioners, as key factors to be considered when modelling, designing, configuring, or managing multi-scale feedback systems within each specific domain. 

\section{Background \& Related Work}
\label{background}

\subsection{Timing in Multi-Scale Systems}

The role of timing 
in multi-scale systems has been explored mostly in either domain-specific ways (e.g., hierarchical smart grids \cite{steghofer2013}, houses \cite{allerdingOC2011} and vehicles \cite{Albus2002-il}) or in generic 
terms (e.g., multi-level design patterns in self-adaptive systems \cite{weyns2013}, 
autonomic systems \cite{kramer2007}, organic computing (OC) \cite{SchmeckOC2011}, self-aware systems \cite{diaconesSAw2017}, and multi-agent systems \cite{Minsky1988}). In both cases, results are difficult to reuse and transfer across domains. 
It is generally considered that lower levels should execute faster than higher ones. While this applies to most systems, the underlying constraints and variants are rarely discussed. Exceptions may also exist depending on desired behaviour (e.g., stock markets may not aim to reach steady state). 

Similar examples can be found in natural system studies. 
In the field of ecology, multi-scale systems are usually nested. 
As macro-properties at higher levels arise from the composition of levels below, it is often taken for granted that higher levels operate slower than lower ones, and that this is necessary for system stability \cite{allen2017hierarchy}. Similarly, research in biology and paleontology that has focused on nested hierarchies assumes that different timescales (with higher levels operating slower than lower ones) are inherent to such systems \cite{valentine1996hierarchies}. Institutional and policy studies, on the other hand, tend to focus on multi-scale systems with exogenous macro-entities (e.g., higher-level bureaucracies send commands to lower ones). 
Here, delays are often described as dysfunctional, as they can lead to policy ineffectiveness (as upper levels send out-dated commands to controlled resources) \cite{brinkman2005cultural} \cite{mangin2018cost} \cite{warren2005hierarchy}. Building on the examples of coral reef formation and power dynamics in Macaque societies, \cite{flack2013timescales} argue that slow variables (at the macro-scale) lead to the adaptation of micro-entities by reducing environmental uncertainty, but that if these variables are too slow, they cannot be detected by micro-entities, leading to a slow variable lock-in. The fact that the slowness of macro-entities allows for them to be perceived as constant by micro-entities is also highlighted by \cite{fairhall2001efficiency} in the context of adaptive neural code, showing how in the vision system of flies adaptation occurs at different timescales,
with longer ones providing a separate information transmission channel. 
\cite{bellman1984} links organism motor functions to primitive language, indicating that macro-properties (or `symbols') allow to delay immediate reactions to external changes, so as to take into account previous experiences and  generate more complex behaviours.

The control systems community has studied timing in multi-scale systems using different terminology, e.g., hierarchical, singularly perturbed, multiloop, nested, and cascade control systems.  Hierarchical control systems are addressed in \cite{Findeisen}, where hierarchies are defined by functions or time horizons of the multilayer configuration. The highest layer necessarily has the longest time horizon to achieve optimal control for the system.  Applications of singular perturbation theory to control systems were reviewed in \cite{Kokotovic}, where systems are decomposed into parts with fast and slow dynamics.  Multiloop \cite{Coffey}, nested \cite{Allwright}, and cascade \cite{Franks} systems generally refer to systems in which multiple feedback loops control variables of importance at different scales.  For example, in aerospace applications the different loops address (from micro- to macro-scale): attitude, attitude rate, and guidance.  In these applications it is generally assumed that higher levels (corresponding to outer loops in the nested system) operate at a slower rate, or that it can be shown in specific situations what the relative rates should be for stability and optimal performance.

\subsection{Coupled Oscillators}
Our two case studies build on existing models of coupled oscillators. Kim et al.'s model \cite{Kim2010} explores the synchronisation of coupled biochemical oscillations in cellular systems. 
Synchronisation is affected by the coupling strength, with two thresholds defining three different behaviours: oscillation without synchronisation, oscillation with synchronisation, and no oscillation. We expand on this model by connecting oscillators in 
multi-scale configurations, and adding micro-macro communication delays 
that affect synchronisation (sec.~\ref{biochemical}). 

For the HCA case, we expand on the model in \cite{diaconescu2018holonic}, where Cellular Automata (CAs) were organised in a multi-scale configuration, generating macro-structures from uniform micro-scale conditions. We expand on this model to explore how different combinations of execution frequencies at various scales affect micro and macro oscillating behaviours (sec.~\ref{HCA}). 

As highlighted in \cite{monsivais2020dynamics}, natural oscillatory processes tend to follow a multi-scale organisation, with macro-scale frequencies affecting micro-scale behaviour. In most oscillators, communication and adaptation are not instantaneous \cite{ramirez2012effects} \cite{ermentrout2009delays}. E.g., biological systems require a minimum interval to transmit information \cite{choi2000synchronization}. The same applies to most artificial systems. Hence, time-related questions are relevant both in the context of oscillator behaviour, 
within a wide range of applications,
and for multi-scale feedback systems, more generally. Several studies discussed the impact of time delays on the behaviour of coupled oscillators (e.g., \cite{choi2000synchronization} \cite{jeong2002time} \cite{ko2004wave}). These studies suggest that time delays can significantly affect system dynamics \cite{earl2003synchronization}.

\section{MSAF Design Pattern \& Time Extensions}
\label{MSAFpattern}

\subsection{Overview of the MSAF Design Pattern}

\begin{figure}
\centerline{\includegraphics[width=6.5cm, height=4cm]{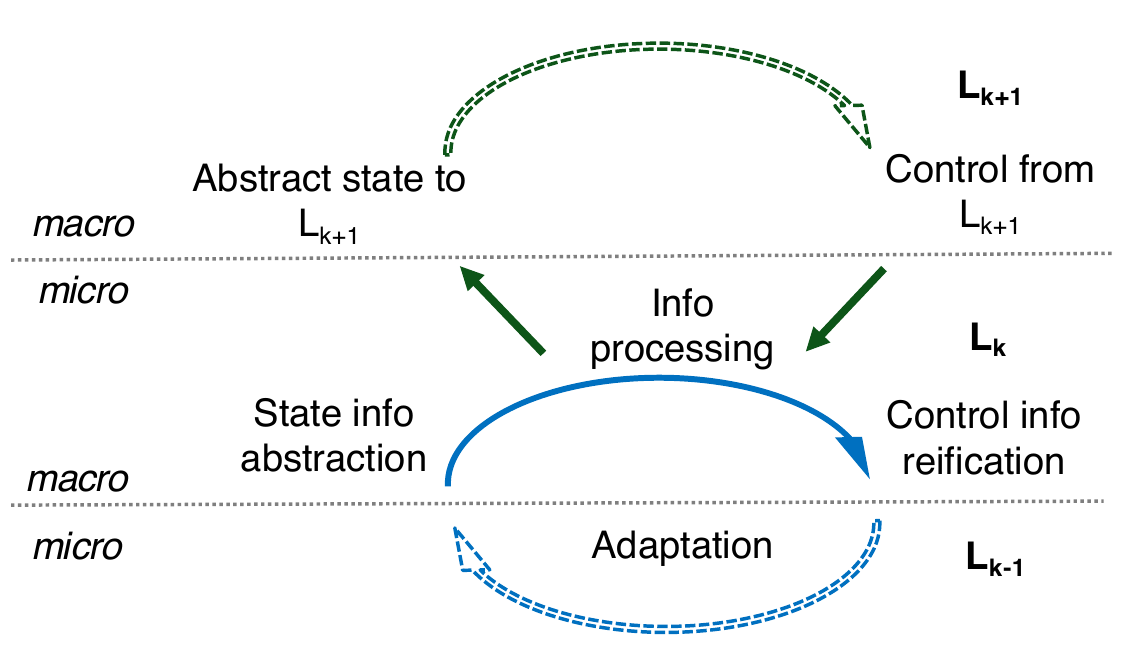}}
\caption{MSAF Feedback Loops: all arrows are information flows; each level $L_k$ is a macro level relative to the one below $L_{k-1}$ (except for bottom level, $L_0$) and a micro level for the one above $L_{k+1}$ (except for top level, $L_{M-1}$)}
\label{fig:msfs-loops}
\end{figure}

The MSAF design pattern \cite{diaconescu2019multi} \cite{diaconescu2021exogenous} models feedback loops in multi-scale systems in terms of information flows 
that merge, split, and cycle through different abstraction levels (Fig.~\ref{fig:msfs-loops}).  Information flows are streams of changes (attached to a material substrate) which can be observed, interpreted, and used for adaptation in line with semantic definitions of information \cite{jablonka2002information}. 

Such information flows merge and aggregate information at increasingly higher abstraction levels (bottom-up), then split and reify information again at 
more detailed levels (top-down), forming multi-scale feedback cycles.

A single feedback loop consists of the following steps: 1) collection and abstraction of state information; 2) information processing (e.g., decision); 3) information reification (control command); and 4) adaptation.  These steps match existing feedback designs in autonomic (MAPE-K) \cite{kephart2003vision} and organic computing 
\cite{SchmeckOC2011}, or feedback control systems \cite{hellerstein2004}.

Extending this design to multiple scales implies adding further 
feedback loops on top of each other. 
This involves two extra steps for connecting feedback loops between levels (in green): a) sending state abstraction of $L_k$ to upper level $L_{k+1}$; and b) receiving control information from $L_{k+1}$, to be used as control input, or goal, in $L_k$'s processing step (2). From $L_k$'s perspective, all upper-level feedbacks can be modelled as a single one (dotted green arrow), at $L_{k+1}$; 
and all lower levels as one adaptation process (dotted blue arrow), at $L_{k-1}$.

Hence, a managed resource (at $L_0$) 
receives feedback controls that merge information from several scales, covering increasingly larger system scopes.
Such multi-scale feedback design helps to control large-scale systems 
by limiting the amount of 
processed information 
at each level and by mixing quick local reactions with slower coordinated responses.

\subsection{Time Considerations in Multi-Scale Feedback Loops}
\label{subsec:ms-timing}

Generalising from feedback systems \cite{hellerstein2004} and control theory \cite{Nise}, we distill several key timing considerations impacting system behaviour: 
communication delay; processing time; adaptation lag; sample time (for digital systems). 
%
To simplify, we merge these into two main timing aspects, applicable to all MSAF steps: 
i) \textit{execution delay} ($\tau$), the step execution duration (including communication and processing); and ii) \textit{execution interval} ($\Delta t$), how often the step executes.
We group MSAF steps (1-3) (abstraction, processing, and control) into a single `management flow' (including inter-level abstraction (a) and control (b) for higher levels), featuring an execution delay $\tau_{mng}$ and interval $\Delta t_{mng}$. The adaptation step (4) also features an execution delay $\tau_{adpt}$ and interval $\Delta t_{adpt}$.

All timing considerations from `classic' feedback control systems apply here. We highlight some of these below 
without aiming for a comprehensive review.  
Delay in the management flow $\tau_{mng}$ implies the risk of providing a control command (output) based on an outdated monitored state (input). It may lead to oscillations, longer settling times, or instability \cite{hellerstein2004}; and decrease reactivity to state disturbances.  
Yet, if $\tau_{mng} << \tau_{adpt}$ there is a risk of overreaction from the management flow, i.e., repeating or exacerbating a control command as it fails to perceive the effects of a previous command. This risk is removed when controls are not `cumulative' (e.g., goal-oriented commands can be repeated with the same effect). 

With respect to execution intervals, the smaller the $\Delta t_{mng}$ (i.e., the management flow executes more often), the more reactive it can be to state changes, while 
again, risking to overreact 
if it executes before previous controls take effect. Overreaction is avoided if $\Delta_{adpt} < \Delta_{mng}$ (also considering delays
); or when controls are merely repeated (without increased amplitudes) and the adaptation flow only executes the last one 
(if $\Delta_{adpt} > \Delta_{mng}$). 
Ideally, the management flow would be fast to execute ($\tau_{mng} \rightarrow 0$) but only execute at intervals large enough to allow for the effects of its commands to take effect in the adaptation flow ($\Delta_{mng} \thicksim \tau_{adpt}$). Other combinations of execution delays and intervals are also viable (domain and application-specific).


The above considerations become more complex when feedback cycles extend across multiple scales, incurring further cross-scale delays and combinations of their relative values. 
%
%
When management flows at different scales execute in parallel (common case), each flow at $L_k$ gets abstract state information from $L_{k-1}$ and control information from $L_{k+1}$, to issue commands back to $L_{k-1}$. For $L_k$, information from $L_{k-1}$ is more recent than from $L_{k+1}$, as the latter would have crossed at least an extra scale. Yet, information from $L_{k+1}$ includes abstractions about broader system scopes (under control levels from $L_{k+1}$ to $L_{M-1}$). This allows $L_k$ managers to coordinate their local actions based on wider system views. 
%
Hence, control information for $L_0$ entities merges information flows from all system scales, with lower-scales information being narrower but more recent (or accurate) and higher-scales information being broader but more outdated.    


Hence, higher-level management flows would have larger delays than lower ones, as it takes more time for their input and output flows to travel to and from $L_0$.  This situation is due to system implementation constraints (i.e., inherent communication and processing delays, at each level), rather than being a desirable system design property.
Still, in case of rapid management relative to adaptation delay ($\tau_{mng}$$<<$$\tau_{adpt}$) it makes sense to execute higher managers less often than lower ones ($\Delta_{mng,k}$$>$$\Delta_{mng,k-1}$), to avoid overreactions or instability.  
However, increasing $\Delta_{mng,k}$  may decrease the system's coordinated responses. 
Typical solutions combine fast, accurate, localised reactions from lower-scales, for avoiding disaster (e.g., reflexes in organisms, obstacle avoidance in autonomous cars) with slower, more context-aware responses, for coordinated behaviour (e.g., strategic planing in organisms, rerouting autonomous cars).      
Various combinations of cross-level execution delays and intervals lead to different system behaviours (macro-properties). We 
focus 
on a few examples illustrated via our two applications (HO and HCA).

\section{Biochemical Oscillator Model}
\label{biochemical}
\subsection{HO Overview}
\begin{figure}
\centerline{\includegraphics[width=3.2in]{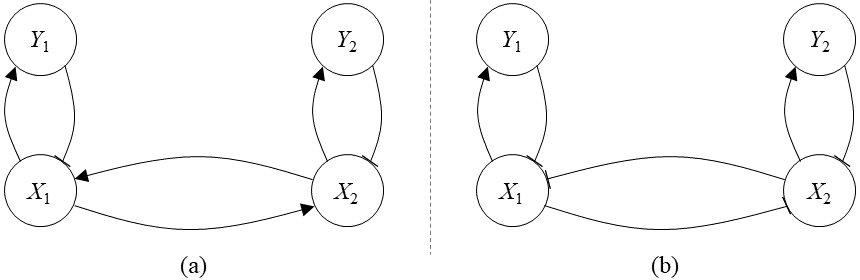}}
\caption{Coupled oscillators with PP coupling (a) and NN coupling (b).}
\label{fig:coupled_oscillator}
\end{figure}

We use the coupled biochemical oscillator model from \cite{Kim2010} which is extended to: i) a flat network of more than two coupled oscillators; and ii) a hierarchy of oscillators (HO).
%
In coupled biochemical oscillators, each oscillator consists of two interacting components $X$ and $Y$, with coupling between the $X$ components as shown in Fig.~\ref{fig:coupled_oscillator}.

\begin{figure}
\centerline{\includegraphics[width=3.2in]{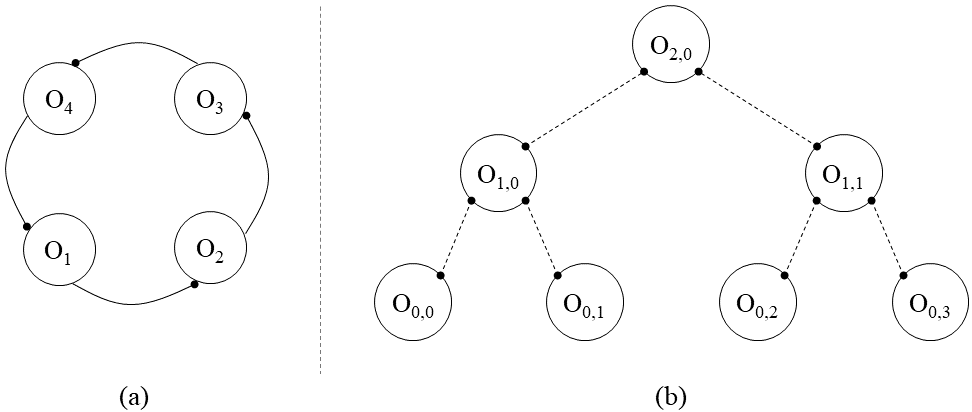}}
\caption{(a) A flat network with $N = 4$ oscillators and (b) a hierarchy of oscillators with $M = 3$ levels and $C = 2$ children per oscillator.  Each oscillator consists of $X$ and $Y$ components.  The small circles indicate either P or N type coupling between the $X$ components of each oscillator.}
\label{fig:oscillator_configurations}
\end{figure}

To accommodate multiple oscillators, we 
generalise 
this model to include more 
oscillators arranged in a 
``flat'' configuration (i.e., where all the oscillators are peers and there is no hierarchy present in the network).  
Fig.~\ref{fig:oscillator_configurations}(a) shows this system with four oscillators, which can have either P or N type coupling.  Each $X_i$ promotes or inhibits $X_{i+1}$ depending respectively on P or N type coupling.

The HO uses a repetitive design, with oscillators being the entities repeating at different levels.  When integrated within a multi-scale structure, these entities differ from their stand-alone forms by taking into account (detailed) state information from the lower levels and (abstract) control information from the upper levels.  The macro-entities are oscillators whose models are modified from \cite{Kim2010}.  Fig.~\ref{fig:oscillator_configurations}(b) shows the structure of the hierarchy of oscillators for three levels with two children per oscillator.  The coupling occurs between the $X$ components of the oscillators between levels.  There is no direct communication between oscillators at a given level.

\subsection{Flat and HO Models}
The flat network of oscillators is modeled by the following differential equations; for $i = 1..N$, where $N$ is the number of oscillators and $i-1$ is within the range $1..N$ (so if $i = 1$, then $i-1 = N$).  It is assumed that all oscillators are coupled with the same strength $F$, type P or N, and time delay $\tau$.
\begin{align}
\begin{split}
\frac{dX_i}{dt} = &\frac{1+PP \left( FX_{i-1}(t-\tau) \right)^3}{1+ \left( FX_{i-1}(t-\tau) \right)^3 + \left( \frac{Y_i(t-2)}{0.5} \right)^3} \ldots \\
&\ldots - 0.5X_i(t) + 0.1
\end{split} \\
\frac{dY_i}{dt} = &\frac{\left( \frac{X_i(t-2)}{0.5} \right)^3}{1+ \left( \frac{X_i(t-2)}{0.5} \right)^3} - 0.5Y_i(t) + 0.1 \label{eqn:Y_flat}
\end{align}

For the HO model, the differential equations for $X$ depend on their level. 
At the highest level, 
oscillators only act as aggregators 
of information from the lower level. 
At the lowest level, 
they only receive feedback from their corresponding macro-entity. 
Oscillators at 
middle levels receive information from above and below. 
While the equation modelling the $Y$ component of all oscillators is analogous to (\ref{eqn:Y_flat}), those for the $X$ components of the bottom, top, and middle levels respectively are:
\begin{align}
\nonumber \frac{dX_{0,i}}{dt} =&\frac{1+PP \left( F_0 X_{1,p_{0,i}}(t-\tau_0) \right)^3}{1+ \left( F_0 X_{1,p_{0,i}}(t-\tau_0) \right)^3 + \left( \frac{Y_{0,i}(t-2)}{0.5} \right)^3} \ldots \\
&\ldots -0.5X_{0,i}(t) + 0.1 \label{eqn:top_X} \\
\nonumber \frac{dX_{K,0}}{dt} = &\frac{1+PP \left( F_{K} \bar{X}_{K,0}(t-\tau_{K}) \right)^3}{1+ \left( F_K \bar{X}_{K,0}(t-\tau_{K}) \right)^3 + \left( \frac{Y_{K,0}(t-2)}{0.5} \right)^3} \ldots \\
&\ldots - 0.5X_{K,0}(t) + 0.1 \\
\nonumber \frac{dX_{m,i}}{dt} = &\frac{1+PP \left( F_m \Gamma_{m,i}(t-\tau_m) \right)^3}{1+ \left( F_m \Gamma_{m,i}(t-\tau_m) \right)^3 + \left( \frac{Y_{m,i}(t-2)}{0.5} \right)^3} \dots \\
&\ldots - 0.5X_{m,i}(t) + 0.1 \label{eqn:middle_X}
\end{align}
where
\begin{align}
\Gamma_{m,i}(\cdot) = &W X_{m+1,p_{m,i}}(\cdot) + (1-W)\bar{X}_{m,i}(\cdot) \label{eqn:gamma}
\end{align}
and $\bar{X}_{m,i}(\cdot)$ denotes the mean $X$ concentration taken over the children of $X_{m,i}$;
%
$m = 0..M-1$, where $M$ is the number of levels; $i = 0..N_m-1$, where $N_m$ is the number of oscillators in level $m$; $p_{m,i}$ is the position of the parent of $X_{m,i}$; $K = M-1$; $F_m$ is the coupling strength; $\tau_m$ is the time delay; and $W \in [0,1]$.  Parameters $F_m$ and $\tau_m$ are constant across a level and $W$ is constant across the system.

In the HO model, the abstracted information is the average concentration of a substance $X$ in micro-entities given by 
$\bar{X}_{m,i}(\cdot)$.  The feedback information that is sent down from macro to micro is the concentration of $X$ in the macro-oscillator, which impacts the concentration of $X$ in the micro-oscillators through (\ref{eqn:top_X}) and (\ref{eqn:middle_X}).  
Function $\Gamma_{m,i}(t-\tau_m)$ in (\ref{eqn:gamma}) communicates both the abstracted information from the lower level and feedback information from the higher level, with $W$ controlling the relative importance of the feedback signal.

\subsection{Simulation Time and Sequence}
The oscillators are continuous time systems described by differential equations, digitized using a numerical solver.  The coupled oscillator systems were simulated in MATLAB using the dde23 function to numerically solve the delay differential equations \cite{Kim2010}.  Each differential equation was solved simultaneously for specified interval of time $[0,T_{end}]$, resulting in a discrete time series of $X$ and $Y$ concentration values.

With respect to timing aspects described in Section \ref{subsec:ms-timing}, there is a micro-to-macro abstraction delay and macro-to-micro feedback delay, both characterised by $\tau_m$, which reflects the time it takes for the concentration of $X$ to be transmitted (i.e., transmission delay, in literature). This means that the abstracted state and feedback is based on old micro-state information.  These semantics simulate communication delays $\tau_{mng}$ for abstracted states, with negligible adaptation delay ($\tau_{adpt}=0$) and continuously executing feedback cycles ($\Delta_{mng}$ and $\Delta_{adpt}$).  Delays at higher levels are always higher than at lower levels due to delay accumulation.

\subsection{Experimental Settings}
A number of system configurations were simulated.  Flat networks had from 4 to 11 oscillators.  HOs had 64 bottom-level oscillators arranged in two ways: $M = 2$ levels, $C = 64$ children; and $M = 4$ levels, $C = 4$ children.  For each configuration a range of parameter values was used: coupling strength $F = 0..8$, time delay $\tau = 1..15$, and coupling (PP, NN).  Each configuration was simulated for 10 runs, with the initial $X$ values of an oscillator 
randomly chosen from the interval (0,1) and the initial $Y$ values 
set to 1.  Values of oscillation frequency, amplitude, and synchronisation time were averaged over the 10 runs.

\subsection{Overall Behaviour}
\begin{figure}
\centerline{\includegraphics[width=3.2in]{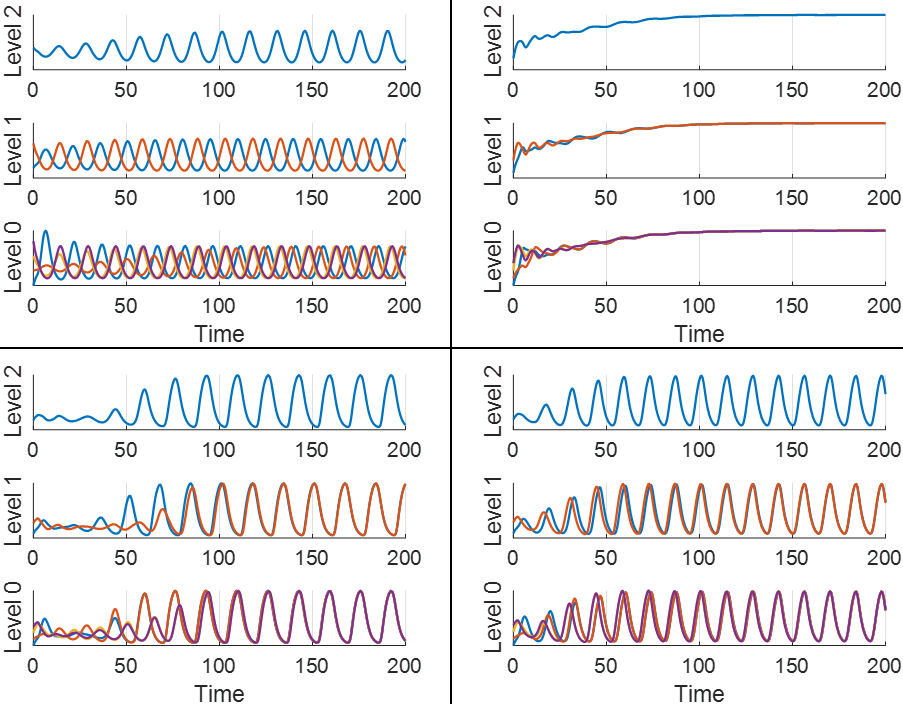}}
\caption{Different types of emergent behaviour for an HO with $M = 3$ levels and $C = 2$ children.  Upper left: unsynchronised oscillation.  Upper right: no oscillation.  Lower left: synchronised oscillation with adjacent levels out of phase.  Lower right: synchronised oscillation with all levels in phase.}
\label{fig:behavior_types}
\end{figure}

The coupled oscillator systems exhibit three basic types of emergent behaviour: unsynchronised oscillation, no oscillation, and synchronised oscillation with all levels in phase.  In the case of the HO, there is an additional type: synchronised oscillation with 
levels 
out of phase.  These behaviours are shown in Fig.~\ref{fig:behavior_types} 
(for $M = 3$ levels and $C = 2$ children).
The three stacked time series plots show the oscillator $X$ concentrations at each level plotted together, with the top level having one oscillator, the middle level having two oscillators, and the bottom level having four oscillators.

\subsection{Experimental Results}
In flat networks, it was found that synchronisation occurred consistently in PP coupled systems having no more than 5 oscillators and for NN coupled systems having no more than 10 oscillators in the network.  
Furthermore, the region of the parameter space that achieved synchronisation was a relatively small subset.  To achieve consistent synchronisation in systems with more than 10 oscillators and for a larger range of $F$ and $\tau$ values, it is necessary to have a hierarchical structure. 

For HO systems, we present two kinds of results relevant to our contribution (with analogs in the HCA model): 
the effect of 
system parameters on (1) generated macro patterns and (2) oscillation periods.  Other results on 
oscillation amplitudes 
and synchronisation settling time are also briefly discussed.

\subsubsection{Impact of time on generated macro patterns}
\begin{figure}
\centerline{\includegraphics[width=3.5in, height=2.2in]{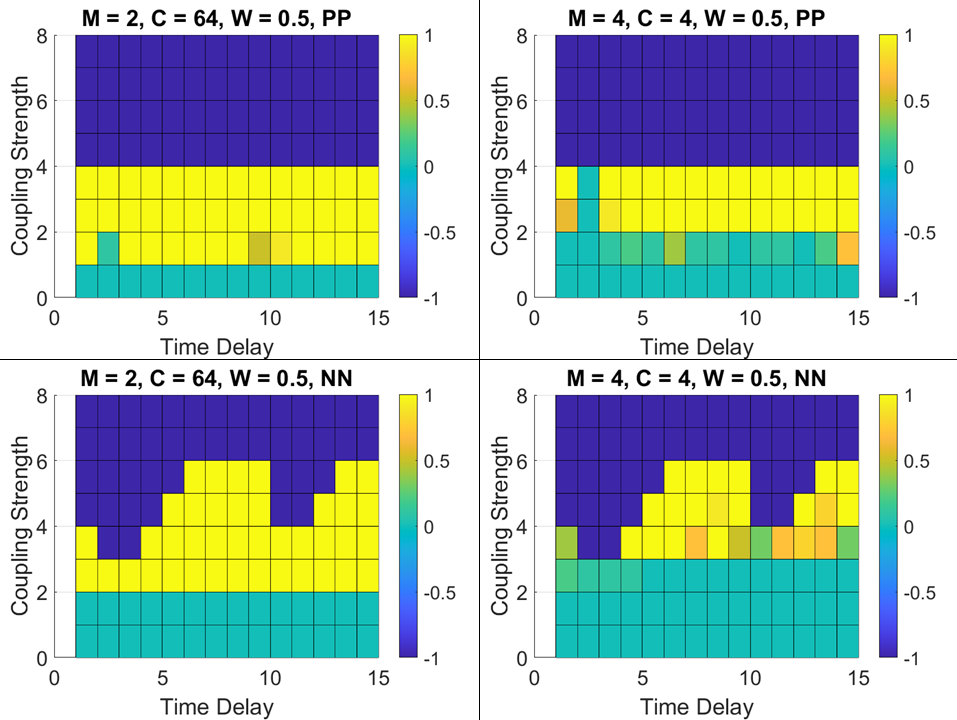}}
\caption{Each plot shows the emergent behaviour for $F = 0..8$ and $\tau = 1..15$. Colors indicate regions in the parameter space where different behaviours occur: Dark blue, no oscillations; yellow, synchronised oscillations (with levels in or out of phase); light blue, unsynchronised oscillation; color gradients from yellow to light blue, synchronisation  only in some runs.}  
\label{fig:oscillator_results}
\end{figure}

In HO systems, Fig.~\ref{fig:oscillator_results} shows the emergent behaviour of the system for all configurations tested.  In all cases, the bottom level synchronised for the middle range values of $F$ (yellow region).  For low $F$ values, 
there was oscillation, but no synchronisation (light blue region). For high $F$ values, 
there were no oscillations (dark blue region). 
The smaller hierarchy ($M = 2$) achieved synchronisation in a larger part of the parameter space.
There are two distinct transition regions: from unsynchronised to synchronised oscillations and from synchronised oscillations to no oscillations.  The transition from synchronised to no oscillations was deterministic. For both PP systems, oscillations only occurred for $0 \le F \le ~3.5$.  This transition was unaffected by time delay.  In contrast, for both NN systems, the transition from oscillations to no oscillations occurred for $F$ between 3 and 6, depending on the value of time delay.  The transition from unsynchronised to synchronised oscillations was stochastic, as indicated by the color transition from light blue to yellow: 
yellow, synchronisation happened in each run; light blue, it did not happen in any run; in-between colors, synchronisation occurred only in some runs, according to the color scale to the right of the plot. 

\subsubsection{Impact of time on oscillation periods}
\begin{figure}
\centerline{\includegraphics[width=3.5in, height=2.3in]{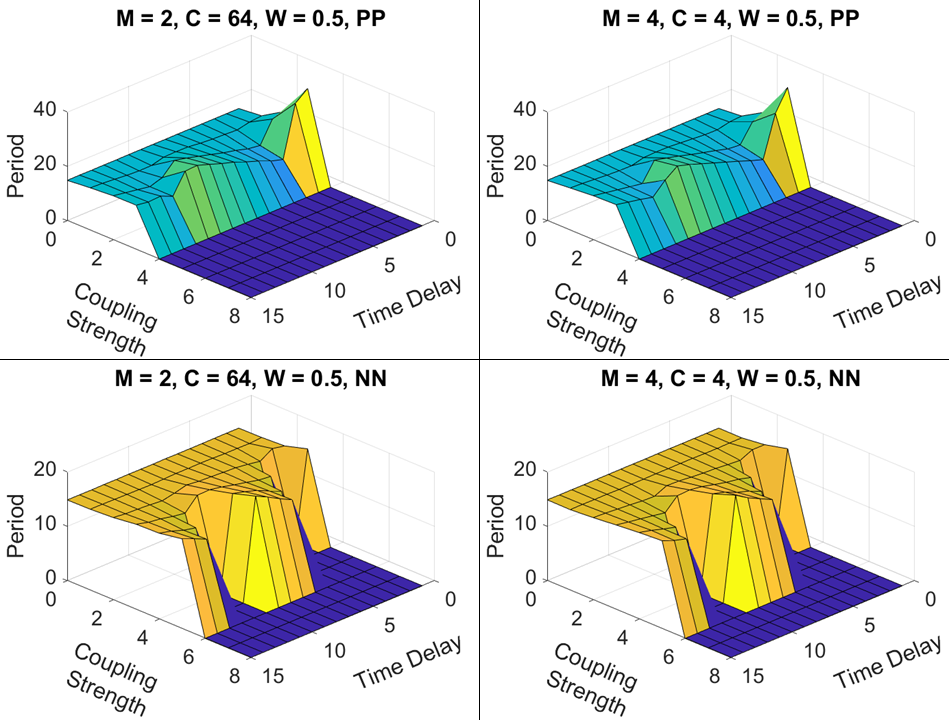}}
\caption{Each plot shows the oscillation period for $F = 0..8$ and $\tau = 1..15$.  Upper left: $M = 2$ levels, $C = 64$ children, PP coupling.  Upper right: $M = 4$ levels, $C = 4$ children, PP coupling.  Lower left: $M = 2$ levels, $C = 64$ children, NN coupling.  Lower right: $M = 4$ levels, $C = 4$ children, NN coupling.  Zero values means there is no oscillation in that region.}
\label{fig:period}
\end{figure}

Fig.~\ref{fig:period} shows how the period of oscillation varies with the coupling strength and time delay.  Time delay has a larger impact on the period for PP coupled systems while the effect is negligible for NN coupled systems.  The number of levels had no effect on the period.

Oscillation amplitude varies 
with 
time delay within the synchronised region, with 
a 
larger impact occurring in NN systems 
(in contrast to the effect on period).  
The number of levels ($M = 2$ and $M = 4$) had a negligible impact on amplitude; but fewer levels lead to faster synchronisation, for both PP and NN coupling. 
The effect of time delay on synchronisation time is more pronounced for 
NN coupling.  Full results are omitted due to space constraints 
(Cf. \href{https://gitlab.telecom-paris.fr/ada.diaconescu/msaf/-/tree/master/acsos21}{ https://gitlab.telecom-paris.fr/ada.diaconescu/msaf (acsos21 directory)}).


\subsection{Discussion}
These results show the effect of time delay on the system's macro patterns and their properties.  For NN coupling $\tau$ affects the type of synchronisation, but not the amplitude.  Conversely for PP coupling, $\tau$ affects the oscillation period, but not the synchronization type.  For both types, the number of levels $M$ affects unsynchronised to synchronised transition, due to the increased time delay caused by larger $M$.  Further these results indicate that HO systems are advantageous compared to flat networks:  (1) HO systems are able to synchronise more oscillators: 
64 oscillators for HO, compared to a maximum of 5 and 10 for flat network (with PP and NN, respectively).  (2) The desired synchronisation behaviour occurs in a larger region of the parameter space as noted by the large yellow regions in Fig.~\ref{fig:behavior_types}.  In contrast, at their maximum size, flat networks achieved synchronisation for a single combination of coupling strength and time delay.  (3) Due to their large synchronised region, HO systems are more robust to parameter variations.  Moreover for NN coupling, a change in one parameter (time delay) can also be compensated by changing another parameter (coupling strength) to achieve synchronisation without affecting the amplitude.

\section{Hierarchical Cellular Automata Case Study}
\label{HCA}

\subsection{HCA Overview}
Cellular Automata (CA) are discrete models where the state of each entity (cell) at $t$ depends on the cell's previous state and on its neighbours' states, at $t$-1. Cells are usually arranged in a grid and their inter-dependency modelled via a rule set. CA, including coupled CA, have been employed to model a wide range of complex systems, including multi-scales \cite{hoekstraCA2021}. 
To analyse timing effects on such multi-scale systems, 
we reuse the Hierarchical Cellular Automata (HCA) simulator in \cite{diaconescu2018holonic}. 
%
It organises multiple CA into several scales (levels). 
Cross-level CA interactions follow the MSAF pattern: 
a) abstract state-information (bottom-up); and b) control commands, or goals (top-down). 
Each CA (except the top) has two rule sets: 
Expansive rules ($R_E$) increase the CA's number of live cells; Regressive rules ($R_R$) decrease them. The control goal from above dictates the CA's active rules (to execute). CAs at different levels have different $R_E$-$R_R$ rule-pairs.

Each CA at a lower level $L_k$ is \textit{mapped} bidirectionally to a single cell of a CA at a higher level $L_{k+1}$. In the bottom-up mapping (a), the entire state of a lower CA is abstracted (based on the percentage of its live cells relative to a threshold $Th_k$) and sets the binary state of its mapped cell in a higher CA. In the top-down mapping (b), the state of each cell in a higher CA controls the rule activation of its mapped lower CA (i.e., sets $R_E$ or $R_R$). These bidirectional interactions form inter-level feedbacks, replicated at successive levels, up to the top (which only executes static rules). Simulations 
are deterministic.

\subsection{HCA Notation \& Inter-level Mapping}

Table \ref{tab:notations} summarises the main HCA concepts and notations (details in \cite{diaconescu2018holonic}). HCA consists of several \textit{levels} ($L_k$), each with one or several CA ($CA_{k,i}$) (Fig.~\ref{fig:3LevelHCA}). 
Each $CA_{k,i}$ at a micro-level is mapped bidirectionally to one cell $C_{k+1,j,s}$ of a $CA_{k+1,j}$ at the macro-level: 
(a) the state abstraction ($AS_{k,i}$) of $CA_{k,i}$ (micro) is set as the state ($CS_{k+1,j,s}$) of its mapped cell $C_{k+1,j,s}$ (macro) (Eq. \ref{eq:abstract-state-transfer} and Eq. \ref{eq:abstract-state-calc}); 
and (b) the control goal ($G_{k,i}$) from the cell state $CS_{k+1,j,s}$ (macro) sets the active rule of its mapped $CA_{k,i}$ (micro) (Eq. \ref{eq:goal-transfer} and Eq. \ref{eq:rule-setting}). 
%
\begin{equation}
\label{eq:abstract-state-transfer}
CS_{k+1,i} \gets AS_{k,i} \;, \;\; i=1..N_k \;, \; map(C_{k+1,i}; CA_{k,i})  
\end{equation}
%
\begin{equation}
\label{eq:abstract-state-calc}
AS_{k,i}=
\begin{cases}
1,& \text{if } \sum_{s=1}^{S_{k,i}} CS_{k,i,s} >= Th_{k}\\
0,& \text{otherwise}
\end{cases}
\end{equation}
\begin{equation}
\label{eq:goal-transfer}
G_{k,i} \gets CS_{k+1,i}, \; \; map(C_{k+1,i}; CA_{k,i})
\end{equation}
%
\begin{equation}
\label{eq:rule-setting}
R_{k,i}=
\begin{cases}
R_{k,E} \; \text{(Expansive rules)},& \text{if } G_{k,i} == 1\\
R_{k,R} \; \text{(Regressive rules)},& \text{if } G_{k,i} == 0
\end{cases}
\end{equation}

\begin{center}
\begin{table}
\begin{tabular}{| p{1.8cm} | p{6.6cm} |}
\hline	
\textbf{Notation} & \textbf{Description} \\ \hline
$L_k$ & Level $k$, with $k=0..M$-$1$, $M$ the N$^\circ$ of HCA levels \\ \hline
$CA_{k,i}$ & \parbox[t]{6.5cm}{Cellular Automata $i$ at level $L_k$, \\$i=0..N_k$-1, $N_k$ the N$^\circ$ of CA at $L_k$} \\ \hline
\parbox[t]{1.3cm}{$CA_{k,i} \Rightarrow \\<state>$}  & \parbox[t]{6.6cm}{$CA_{k,i}$ converges to steady state $<state>$: either $O_P$ (oscillate with period $P$) or $S_X$ (stuck with $X$ live cells)} \\ \hline
$CS_{k,i,s}$ & \parbox[t]{6.6cm}{State of Cell $C_{k,i,s}$ of $CA_{k,i}$ ($S_{k,i}$ cells), s=1..$S_{k,i}$\\ $CS_{k,i,s} \in \{0,1\}$, $0 \equiv$ false/dead, $1 \equiv$ true/live}  \\ \hline 
$AS_{k,i}$ & Abstract State of $CA_{k,i}$, $AS_{k,i} \in \{0,1\}$\\ \hline
$Th_{k}$ & Threshold for calculating Abstract States of $CA_{k,i}$ at $L_k$\\ \hline
$G_{k,i}$ & Goal of $CA_{k,i}$; $G_{k,i} \in \{0,1\}$\\ \hline
$R_{k,i}$ & Active Rules (executing) of automaton $CA_{k,i}$ \\ \hline
\parbox[t]{1.8cm}{$map(C_{k,i,s};\\ CA_{k-1,j})$} & Mapping between cell $C_{k,i,s}$ and automaton $CA_{k-1,j}$; implies transfer of abstract state (up) \& cntrl. goal (down)\\ \hline
$Fq_k$ & Activation frequency of level $L_k$-- the number of activations of $L_k$ after which $CA_{k,i}$ actually execute $R_{k,i}$.\\ \hline
$Fq=...$ & \parbox[t]{6cm}{Activation \textit{frequency pattern} across HCA levels; \\ E.g., $Fq$= 1-2-3 means that $Fq_0$=1, $Fq_1$=2, $Fq_2$=3}\\ \hline
\end{tabular}
\caption{Main HCA Concepts and Notations}
\label{tab:notations}
\end{table}
\end{center}

\subsection{Simulation Time \& Sequence}
 
 A HCA simulation proceeds in discrete cycles, each one executing all levels successively, from bottom $L_0$ to top $L_{M-1}$. A cycle consists of $M$ discrete steps $t_k$ ($k$=0..M-1), each one executing all CAs at a corresponding level $L_k$.
 Each $CA_{k,i}$ in an active level $L_k$: \textbf{i) exchanges information with its macro-CA}; (sends $AS_{k,i}$, Eq. \ref{eq:abstract-state-calc}; gets $G_{k,i}$, Eq. \ref{eq:goal-transfer}); 
 \textbf{ii) sets its active rules} ($R_{k,i}$) \textbf{depending on its goal} ($G_{k,i}$, Eq. \ref{eq:rule-setting}); and, \textbf{iii) steps} (executes its active rules). 
 As exceptions, $CA_{0,i}$ (bottom) do not get abstracted states from below, using their previous state instead; and $CA_{M-1}$ (top) do not get goals from above, using a static rule. 
 During a step $t_k$, all CAs at $L_k$ execute in parallel; the step ends when all $CA_{k,i}$ have finished executing.
 
 With respect to timing aspects in subsec.~\ref{subsec:ms-timing}, HCA considers state abstraction delays as negligible; and controls incurring delays of one cycle between each two levels (i.e., $\tau_{mng,k}$=1 cycle). Hence, abstract state is always up to date (i.e., travels across all levels in one cycle) yet controls take $M$ steps to arrive from top to bottom. Adaptation delay is also negligible ($\tau_{adpt}$=0).
 Each level has an activation frequency ($Fq_k$) (i.e.,  execution interval $\Delta_{mng,k}$): $Fq_k$=$d$ means that $L_k$ only activates at every $d$ cycles. 
 %
 Finally, control commands $G_k$ are cumulative (repeating them exacerbates the effect), yet do not increase their values if inactive micro-CAs ignore them.

 \subsection{Experimental Settings}
\begin{figure}
    \centering
    \includegraphics[width=0.4\textwidth, height=3cm]{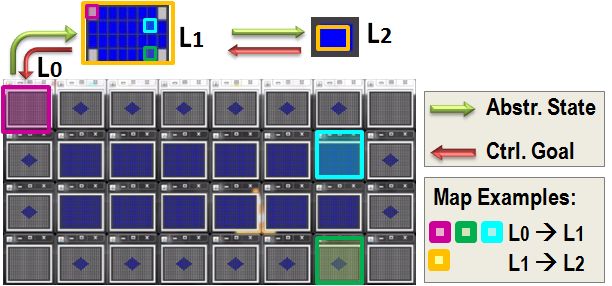}
    \caption{3-level HCA, with 3 differentiated states at $L_0$: 4x Corners $CA_{Cr}$ (purple), 16x Borders $CA_{Bo}$ (green), 12x Core $CA_{Co}$ (cyan)}
    \label{fig:3LevelHCA}
\end{figure}

 We set-up a three-level HCA: $L_0$ (bottom), $L_1$ (middle) and $L_2$ (top), Fig. \ref{fig:3LevelHCA}. $L_0$ has 32 CAs (4x8 matrix), of 441 (21x21) cells each. This maps to a 32 (4x8) cells CA at $L_1$; which maps to a one-cell CA at $L_2$. 
 To simplify HCA behaviour and analysis, we only experiment here with \textit{inversible rule-pairs}: from any CA state, executing $R_E$ and $R_R$ (or $R_R$ and $R_E$) leads to the same state. 
 Non-inversible rules 
 were exemplified in \cite{diaconescu2018holonic}; results presented here do not apply to these.
 
All experiments start with $CA_{k,0}$ in the same initial state, executing $R_{0,E}$; and $CA_1$ and $CA_2$ in dead state (sending $G$=1 control goals until first changing to live states). Experiments vary in configurations for: the two thresholds ($Th_0$ and $Th_1$) for calculating abstract states for $L_1$ and $L_2$ (Eq. \ref{eq:abstract-state-calc}); and the three activation frequencies ($Fq_0$, $Fq_1$ and $Fq_2$), setting the delay between subsequent level activations (for $L_0$, $L_1$ \& $L_2$). 

\begin{figure}
    \centering
    \includegraphics[width=0.48\textwidth, height=1cm]{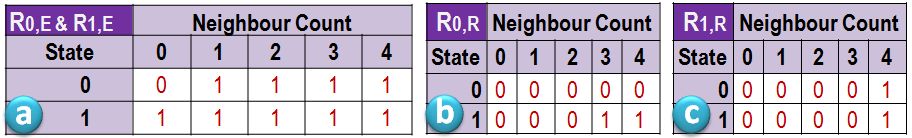}
    \caption{Diamond Rules: a) $R_E$ for $L_0$ \&$L_1$; b) $R_R$ for $L_0$; (c) $R_R$ for $L_1$}
    \label{fig:diamond-rules}
\end{figure}

\begin{figure}
    \centering
    \includegraphics[width=0.48\textwidth, height=2.5cm]{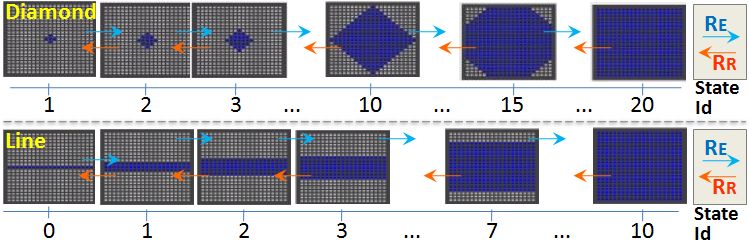}
    \caption{21x21 CA States: \textit{expand} left-to-right, $R_E$; \textit{regress} right-to-left, $R_R$ 
    \textbf{Diamond (top)}: generates 20 states, with N$^{\circ}$ of live cells: 1, 5, 13, 25, 41, 61, 85, 113, 145, 181, 221, 261, 297, 329, 357, 381, 401, 417, 429, 437, 441;
    \textbf{Line (bottom):} 11states: 21, 63, 105, 147, 189, 231, 273, 315, 357, 399, 441}
    \label{fig:diamond-states}
\end{figure}

To show the rule-independence of our results, we tested two inversible rule-pairs at $L_0$: 1) \textbf{Diamond} (Fig. \ref{fig:diamond-rules}: (a) $R_E$ \& (b) $R_R$; non-toroidal), with $CA_{0,i}$ initialised with \textit{one central live cell}, generating an expanding and regressing diamond shape (Fig. \ref{fig:diamond-states}-top); and 2) \textbf{Line} ($R_E$: cell `dead' $\rightarrow$ `live' if at least one live neighbour, and $R_R$: cell `live' $\rightarrow$ `dead' when less than 4 live neighbours; toroidal), with $CA_{0,i}$ initialised with \textit{a central horizontal line of live cells}, generating a vertically-expanding and -retracting rectangle (Fig. \ref{fig:diamond-states}-bottom).

\begin{figure}
    \centering
    \includegraphics[width=0.45\textwidth, height=1.0cm]{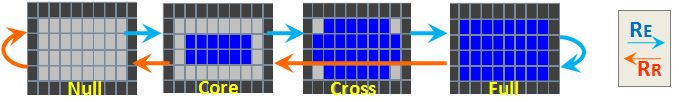}
    \caption{$L_1$ States: expand left-to-right ($R_E$): Null $\rightarrow$ Core $\rightarrow$ Cross $\rightarrow$ Full; regress right-to-left ($R_R$): Null $\leftarrow$ Core $\leftarrow$ Full (skip Cross, non-inversible)}
    \label{fig:L1-states}
\end{figure}

For $CA_1$ we always use a non-inversible rule-pair (Fig. \ref{fig:diamond-rules}: (a) $R_E$ \& (c) $R_R$), which generates four possible states (Fig. \ref{fig:L1-states}): \textit{Null} (0 live cells); \textit{Core} (12 live cells); \textit{Cross} (28 live cells); \textit{Full} (32 live cells). $L_2$ has a single rule-set, inversing the current state (i.e., live $\rightarrow$ dead; dead $\rightarrow$ live).

We define three experimental sets, with $Th_0$=0.1 (i.e., $10\%$ live cells) and $Th_1$ $\in$ \{$0.3$, $0.7$, $0.9$\}. These three values fall in-between the four $CA_1$ states; 
all other values 
are redundant (i.e., same results). Within each 
set, we run tests with varying activation frequencies: $Fq_0$=$1..2$, $Fq_1$=$1..5$, $Fq_2$=$1..5$. 
A test with e.g., $Fq$=1-3-5 means that $Fq_0$=1, $Fq_1$=3, $Fq_3$=5. 
This means about 300 tests (2 rules x 3 $Th_1$ vals. x 50 $Fq$ vals.).

\subsection{Overall Behaviour}

A finite CA can only converge ($\Rightarrow$) to three behaviours: i) \textit{dead} ($S_0$), all cells set to $0$; ii) \textit{live-stuck} ($S_X$), blocked in a state with $X$ live cells (set to $1$); iii) \textit{oscillating} ($O_P$), cycling through a set of states, with the state sequence repeating every $P$ steps. 
At $L_0$, a $CA_{0,i}$'s behaviour depends on the goal pattern received from $L_1$ (i.e., $1$ \& $0$ sequence activating $R_E$ \& $R_R$). If a goal pattern has more $0$s than $1$s, activating  $R_R$ more than $R_E$, then $CA_{0,i}$$\Rightarrow$$S_0$. If $R_E$ activates more than $R_R$, then $CA_{0,i}$$\Rightarrow$$S_{441}$. 
For `balanced' $R_E$ \&$R_R$ patterns, $CA_{0,i}$$\Rightarrow$$O_P$. 

Superposing goal patterns from $CA_1$'s four states differentiates $CA_{0,i}$ into a maximum of three groups (Fig. \ref{fig:3LevelHCA}): 1) Core $CA_{0,Co}$, the 12 $CA_{0,i}$ at the core of $L_0$'s 4x8 matrix; mapped to the 12 live cells in $CA_1$'s Core state; 2) Corner $CA_{0-Cr}$, the 4 $CA_{0,i}$ at the corners of $L_0$'s matrix; mapped to the 4 dead cells in $CA_1$'s Border state; and 3) Border $CA_{0-Bo}$, 16 remaining $CA_{0,i}$ on the borders of $L_0$'s matrix (no corners).   

In brief, $CA_{0,i}$ have an expanding or regressing tendency (i.e., growing or shrinking N$^{\circ}$ of live cells) depending on the active rule set, $R_E$ or $R_R$, respectively. When crossing $Th_0$, this tendency is propagated (and accentuated) upwards through $CA_1$. When crossing $Th_1$, it reaches $CA_2$, which inverses it. The inverse tendency is propagated downwards, back to $CA_{0,i}$, which crosses $Th_0$ the other way. The propagation process is repeated upwards with the opposite tendency, then inverses again at $CA_2$. This creates an expansion-regression oscillation across levels. 
Because $CA_1$ executes its own rule-pair ($R_{1,E}$-$R_{1,R}$), $CA_{0,i}$ differentiate, following different behaviours and converging to different states (e.g., 3 $CA_{0,i}$ states in Fig. \ref{fig:3LevelHCA}).          

\subsection{Experimental Results}

We present two main kinds of results, relevant to our contribution. 
Firstly, we show how different activation frequencies 
constrain the possible oscillation periods $P$ that may occur at HCA levels. We also note that many frequency combinations generate the same oscillation period $P$ (macro-property), though not necessarily through the same state set. 
%
Secondly, we show how different activation frequencies lead to different macro-patterns amongst $CA_{0,i}$, i.e., whether $CA_{0,Co}$, $CA_{0,Bo}$, $CA_{0,Cr}$ converge to $O_P$, $S_0$ or $S_X$. 
The full results set is available from \href{https://gitlab.telecom-paris.fr/ada.diaconescu/msaf/-/tree/master/acsos21}{ https://gitlab.telecom-paris.fr/ada.diaconescu/msaf (acsos21 directory)}. 

\subsubsection{Impact of time on oscillation periods}

\begin{figure}
    \centering
    \includegraphics[width=0.35\textwidth, height=4.5cm]{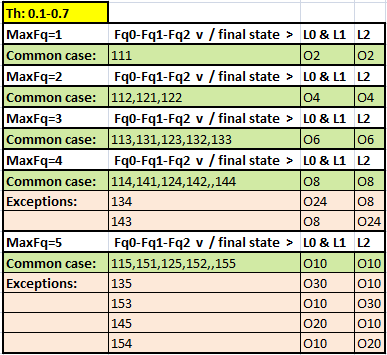}
    \caption{Diamond oscillations: $Th_0$=0.1, $Th_1$=0.7, $Fq_0$=1, $Fq_1$\&$Fq_2$=1..5}
    \label{fig:1-7-freq-results}
\end{figure}

Fig. \ref{fig:1-7-freq-results} summarises results for \textbf{Diamond rules}, with $Th_0=0.1$, $Th_1=0.7$, $Fq_0=1$; and $Fq_1$ \& $Fq_2$ varying between $1$ and $5$. At $L_0$, some $CA_{0,i}$ oscillate (same $O_P$) and some end in a static state ($S_0$ or $S_{441}$). To simplify, we only show here the $O_P$ value for $CA_{0,i}$s that do oscillate; and discuss differentiated $CA_{0,i}$s (i.e., macro-patterns) in the next subsection. Equivalent results were obtained for $Fq_0$=2, in terms of obtained $O_P$ types.

Results show clear correlations between activation frequencies ($Fq$ pattern) and ensuing $O_P$s. We generalise these via unifying formulae (Eq. \ref{eq:P0-1} \& \ref{eq:P2}), deduced from empirical analysis. 
We note that in `common' cases (green in Fig. \ref{fig:1-7-freq-results}), $P$ is the double of the maximum frequency among levels (Eq. \ref{eq:P0-1}-upper for $P_0$ \& $P_1$ and Eq. \ref{eq:P2}-upper for $P_2$).  
E.g., 
for $Fq \in$\{1-3-1, 1-3-2, 1-3-3\}, we get oscillations with $P=6$ ($O_6$) at all levels.   
In `exceptional' cases (red in Fig. \ref{fig:1-7-freq-results}), $P$ is a multiple of the lowest common multiplier (LCM) of all frequencies (Eq. \ref{eq:P0-1}-lower for $P_0$ \& $P_1$ and Eq. \ref{eq:P2}-lower for $P_2$). 
In such cases, we have parameters $a=2$ and $b=2$ in Eq. \ref{eq:P0-1} \& \ref{eq:P2}. E.g., $Fq$=1-3-4 gives $P_0$=$P_1$=24 (LCM=12, b=2) and $P_2$=8; $Fq$=1-4-3 gives $P_0$=$P_1$=8 and $P_2$=24. 
Intuitively, it makes sense for $P_k$ to be some multiple of $Fq_k$, where this multiple may depend on the other $Fq_l$ ($l \neq k$).
\begin{equation}
\label{eq:P0-1}
\begin{split}
P_0=P_1=
\begin{cases}
a*Fq_1,\; \text{if } Fq_1 \geqslant Fq_2 \\
b*LCM(Fq_0,Fq_1,Fq_2),\; 
\text{if } Fq_1 < Fq_2
\end{cases}\\
a, b \in \mathbb{N}_{>0} \setminus \{1\}
\end{split}
\end{equation}
%
\begin{equation}
\begin{split}
    \label{eq:P2}
    P_2=
    \begin{cases}
    a*Fq_2, \; \text{if } Fq_2 \geqslant Fq_1 \\
    b*LCM(Fq_0,Fq_1,Fq_2),\; \text{if } Fq_2 < Fq_1
    \end{cases}\\
    a, b \in \mathbb{N}_{>0} \setminus \{1\}
\end{split}
\end{equation}

Similar results were obtained when increasing $Th_1$ to $0.9$.  
The main difference was that for $Fq$=1-3-4, we obtained $O_{12}$ at all levels, hence $a$=$b$=$1$; rather than $O_{24}$-$O_{24}$-$O_{8}$ as when $Th_1$=$0.7$. 
Results for $Th_1$=$0.3$ were also similar, the only difference occurring for $Fq \in $ \{1-4-5, 1-2-3\}, where the HCA $\Rightarrow$ $S_0$. 
Using \textbf{Line rules} produced equivalent 
$O_{k,P}$ types, when testing the same configuration ranges. 
A difference here is that the toroidal configuration means that a $CA_{0,i}$ that grows ($R_E$) to all live cells can no longer regress ($R_R$), hence staying in $S_{441}$. E.g., for $Fq_0$=1, $Fq_1$=5, $Fq_2$=1..5, we have $CA_{0,i}$$\Rightarrow$$S_{441}$, for all threshold combinations tested. 

\subsubsection{Impact of time on generated macro patterns}
At $L_0$, macro-patterns occur as different behaviours (e.g., $S_0$, $S_X$, $O_P)$ in the three $CA_{0,i}$ groups: $CA_{0,Cr}$ (corners), $CA_{0,Bo}$ (borders), and $CA_{0,Co}$ (core). Different $Fq$ patterns generate different macro-patterns ($CA_{0,Cr-Bo-Co}$). 
%
E.g., for \textbf{Diamond rules} with $Th_0$=0.1, $Th_1$=0.9, we test all frequency combinations having at least one $Fq_k$=3 (i.e., 19 cases) to obtain $O_6$ (or multiples) at $L_0$. This distinguishes three configuration sets, producing three macro-patterns: i) for $Fq \in$ \{1-1-3, 2-1-3, 3-1-2, 3-1-3\} (4 cases), we get $CA_{0,Cr-Bo-Co} \Rightarrow$ \underline{$S_0$-$O_6$-$S_{441}$} (i.e., $CA_{0,Cr} \Rightarrow S_0$, $CA_{0,Bo} \Rightarrow O_6$, $CA_{0,Co} \Rightarrow S_{441}$); ii) for Fq $\in$ \{3-2-1, 3-2-2\} (2 cases), $CA_{0,Cr-Bo-Co} \Rightarrow$ \underline{$O_{12}$-$O_{12}$-$S_{441}$}; iii) for the other combinations (13 cases), $CA_{0,Cr-Bo-Co} \Rightarrow$ \underline{$O_6$-$O_6$-$S_{441}$}.

Similarly, for \textbf{Line rules} with e.g., $Th_0$=0.1, $Th_1$=0.3 and $Fq$ patterns with at least one $Fq_k$=3 (19 cases), we obtain four configuration sets producing four macro-patterns: i) for $Fq \in \{$1-1-3, 3-1-1, 3-3-1, 3-3-2, 3-3-3$\}$ (5 cases), we get $CA_{0,Cr-Bo-Co}$ $\Rightarrow$\underline{$S_0$-$S_0$-$O_6$}; 
ii) for $Fq$=2-1-3 (1 case), $CA_{0,Cr-Bo-Co}$ $\Rightarrow$ \underline{$S_0$-$S_0$-$S_0$}; iii) for $Fq \in \{$3-1-2, 3-1-3, 3-2-1, 3-2-2, 3-2-3$\}$, $CA_{0,Cr-Bo-Co} \Rightarrow$ \underline{$S_0$-$O_6$-$S_{441}$}; and iv) for the rest (8 cases), $CA_{0,Cr-Bo-Co} \Rightarrow$ \underline{$S_0$-$S_0$-$S_{441}$}.

\subsection{Discussion}

Results on the impact of activation frequencies on the resulting behaviour show how oscillation periods $P$ can be controlled via timing adjustments. Interestingly, $P_k$ only depends on  cross-level activation frequencies (Eqs. \ref{eq:P0-1} \& \ref{eq:P2}), while the actual (inversible) rule-pairs and threshold configurations only impact the state set that such oscillations cycle through; and the $CA_{0,i}$ macro-pattern.  
Also notably, a wide range of frequency combinations lead to similar oscillation behaviour, e.g., all combinations with a maximum frequency of 2 lead to $O_4$ forms; maximum $Fq$ of 3 lead to $O_6$ forms (and sometimes multiples, e.g., $O_{12}$); maximum of $Fq$=4 to $O_8$ (and multiples, e.g., $O_{12}$, $O_{24}$); and maximum of $Fq$=5 to $O_{10}$ (and multiples, e.g., $O_{20}$, $O_{30}$). These are important properties for obtaining generic oscillations ($O_P$), while being robust to certain disturbances (e.g., in thresholds or frequencies).      

Results on $CA_{0,i}$ macro-pattern formation show how varying frequencies can lead to equivalent oscillation periods at $L_0$ (as above) yet occurring via different $CA_{0,i}$ groups. Hence, activation frequencies become configuration parameters for shifting overall system behaviour (i.e., macro-pattern productions). As above, different frequency configurations can lead to the same macro-pattern, 
which can enhance robustness.

\section{Discussion, Conclusions \& Perspectives}

This paper aimed to narrow the gap between highly general statements and domain-specific theory about timing in multi-scale feedback systems. It highlighted cross-domain timing aspects, e.g., time delays and execution intervals, and their (combined) impacts on resulting system behaviour (macro-properties). Some of these phenomena were illustrated via two multi-scale oscillator simulators (hierarchical biochemical oscillators (HO) and cellular automata (HCA)) which are both generic and applicable to various domains. 

Experimental results from both examples show how timing confers a configuration parameter just as powerful as any other variable. Changing delays or execution intervals, in various cross-scale combinations, generates different outcomes, e.g., synchronisation type in HO; oscillation period and differentiation pattern in HCA. Several time configuration regions produce equivalent macro-behaviours, possibly improving system robustness to time disturbances. 
This may benefit applications that require rapid behavioural plasticity, without re-learning or reconfiguring other parameters (e.g., neuro-modulated artificial neural networks \cite{neuromodulationPLOSone2020}). 
This contribution sets a basis for developing a comprehensive theory of timing in multi-scale feedback systems, helping practitioners to transfer and apply key insights across specific domains.


\printbibliography

\end{document}